\documentclass[12pt]{article}
\newcommand{\be}{\begin{equation}}
\newcommand{\ee}{\end{equation}}
\newcommand{\beas}{\begin{eqnarray*}}
\newcommand{\eeas}{\end{eqnarray*}}
\newcommand{\bea}{\begin{eqnarray}}
\newcommand{\eea}{\end{eqnarray}}
\newcommand{\ba}{\begin{array}}
\newcommand{\ea}{\end{array}}

\newcommand{\si}{\sigma}

\begin{document} 
%\begin{titlepage}
\title{{
\bf\large  
GRAVISCALAR DARK MATTER AND SMOOTH GALAXY HALOS 
%Graviscalar dark matter and smooth galaxy halos 
%%%%%%%%%% Running Head %%%%%%%%%%%%%
% Graviscalar dark matter
%%%%%%%%%%%%%%%%%%%%%%%%%%%%%%%%%%%%%
}}
\author{Yu.\ F.\ Pirogov
%\thanks{}
\\
%\it Theory Division, 
\it Institute for High Energy Physics,  Protvino, 
\it 142281 Moscow Region, Russia
\\ \it pirogov@ihep.ru
}
\date{}
\maketitle
%\abstract
{\noindent 
In the framework of the unimodular metagravity, with the scalar
graviton/graviscalar dark matter, a regular anomalous one-parameter
solution to the static spherically symmetric  metagravity equations in
empty space is found. The solution presents a smooth graviscalar halo,
with a finite central density profile, qualitatively reproducing the
asymptotically flat rotation curves of galaxies. To refine the
description studying the axisymmetric case in the presence of luminous
matter is in order.
\\[1ex]
{\it Keywords}\/: Unimodular metagravity; graviscalar dark
matter; galaxy halos;  density profiles; 
flat rotation curves.\\[1ex]
PACS Nos.: 04.90.+e, 95.35.+d 
}

\section*{1. Introduction}

General Relativity (GR)  is well-known to be  a theory of the
massless tensor graviton, with two physical degrees of freedom
residing in the metric field. At that,  the unphysical metric
components are restrained due to the general covariance (GC). In this
respect, the unimodular covariance (UC) is a viable alternative to GC.
Namely, it has been shown that
UC is necessary and sufficient to retain two transverse components
for a massless tensor field, with  GC being thus excessive to this
purpose.$^1$ This may {\it de facto} justify  all the GC violating
alternatives to GR, possessing the residual UC. {\it Viz.}, they
may be considered as  the theories of the massless tensor graviton
with the different realizations of a metric component corresponding
to a dilaton. Retaining this component  but making it unphysical  due
to GC we would arrive at GR. Respectively, two routs  to
go beyond GR, with the residual UC, are envisaged. {\em (i)}~To
eliminate such a component  from the metric {\it ab initio} by means
of a unimodularity condition. This would bar the local scale
transformations, with the
local measure becoming an absolute element.$^{2}$ This is the
so-called unimodular relativity/gravity.
(For a recent discussion, see, {\it e.g.}, Refs.~3 -- 4.)  The
cosmological constant emerges here as an integration constant, instead
of a Lagrangian parameter in GR.
Thereupon, one hopes to naturally explain the cosmological
constant being tiny (the long-standing naturalness problem). 
Furthermore, extending the unimodular relativity/gravity by
an exactly massless dilaton one can try to explain the hierarchy
problem in SM unified with gravity and simultaneously  solve the
so-called dark energy problem.$^5$  {\em (ii)}~To convert the
aforesaid unphysical
(but still ``harmless'')  component to the physical one  by  adding to
the GR Lagrangian a GC violating term with a derivative of the metric.
At that, a (massive) dilaton would arise as a part of the metric
field.$^6$  Reflecting GC violation, such a rout originally implies
(a class of) the distinguished, ``canonical'' coordinates, with the
restricted (unimodular) group of the admitted transformations
(thereof, the so-called restricted relativity/gravity). 

In an earlier paper, we put forward a hypothesis that  GC violation
with the metric derivative terms  may serve as a {\it raison d'etre}
for appearance in the Universe of the dark matter (DM) of
gravitational origin.$^7$  The
reason is that under such a GC violation the energy-momentum
tensor of the ordinary matter alone ceases to be covariantly
conserved. This non-conservation can be compensated by equivalently
treating the additional terms in the gravity equations as an
energy-momentum
tensor for the additional gravity degrees of freedom (the extra
``gravitons'') and associating the latter ones with DM. The metric
itself serves thus as a resource of DM.  As a simplest realization of
this approach, the residual UC, with  the local scale covariance
alone being violated, was imposed. In this case, the  metric comprises
just one extra physical degree of freedom, a (massive) scalar
graviton/graviscalar besides the (massless) tensor graviton.
Possessing UC and containing the extra graviton, such a
theory may be called the ``unimodular  metagravity''. 
By introducing a non-dynamical scalar density we put the theory to the
arbitrary observer's coordinates, beyond the canonical ones. 
In a subsequent paper,  the graviscalar field was taken as an
independent variable substituting a metric component in the desired
observer's coordinates.$^8$   This
allowed us to straightforwardly  confront the unimodular metagravity
with GR in the presence of an ordinary scalar field.  
More particularly, an exact ``normal'' solution to the static
spherically symmetric metagravity equations  in the empty, but for a
singular point, space
was written down. The solution is singular in the center and
presents the  black holes filled with graviscalars. It implies the
``normal'' rotation curves (RC's), {\it i.e.}, those declining
asymptotically with distance according to the Newton~law. 

In the present paper, a regular ``anomalous'', missing in GR, solution
to the static spherically symmetric metagravity equations in empty
space is studied. The solution naturally results in the ``anomalous'',
asymptotically flat RC's. It presents a smooth halo as a
coherent state of the graviscalar field in the vacuum. 
Treated in terms of DM the halo possesses a finite central density
profile reproducing qualitatively the contribution to the galaxy RC's
due to DM. The way to refine the description of the galaxy halos, as
composed of the graviscalar DM,  is finally indicated.

\section*{2. Anomalous Vacuum Solution}

\paragraph{Unimodular metagravity}

In the framework of the effective field theory of metric, the
 Lagrangian of the  unimodular metagravity looks most generally
like:$^{7,8}$
\be\label{Lag}
L= L_g +L_h +L_{m} + L_{gh}+L_{mh},
\ee
where the graviton and  graviscalar Lagrangians  $L_g$
and $L_h$, respectively, are as follows:
\bea\label{Lag'}
L_g&=&- \Big(\frac{\kappa_g^2}{2}R+\Lambda\Big), \\
L_h &=&  \frac{1}{2}  \partial \chi\cdot\partial \chi - V_h(\chi),
\eea
with  $\chi$ being the graviscalar field.\footnote[1]{The graviscalar
being a kind of a ``hidden'' particle, the related quantities are
endowed with a subscript~h.}   In the above, 
$\kappa_g=1/(8\pi G)^{1/2} $ is  the GR mass scale, with
$G$ standing for the Newton's constant,
$R$ is the Ricci scalar, $\Lambda$ is the
cosmological constant, $V_h$ is the
graviscalar potential and $\partial \chi\cdot\partial
\chi=g^{\mu\nu}\partial_\mu \chi \partial_\nu \chi $, with
$g_{\mu\nu}$ being the metric.  
The  Lagrangian $L_{m}$ for an ordinary matter has some conventional
form. In the minimal metagravity, we consider, $L_{gh}=L_{mh}=0$.

The peculiarity of the  graviscalar compared to an ordinary scalar
is that the former is not independent of the metric,
{\it viz.},
\be\label{chi}
\chi=\frac{\kappa_h}{2} \ln \frac{g}{g_h}, 
\ee
where $g=\mbox{det\,} g_{\mu\nu}$ and  $g_h$ is a non-dynamical
scalar density of the same weight as $g$. 
In the canonical coordinates, we have $g_h=-1$.
The density $g_h$ makes $\chi$ a GC scalar and allows to bring the
theory to the arbitrary observer's coordinates.  The
parameter $\kappa_h$ stands for a  unimodular metagravity  mass scale 
additional to  the GR $\kappa_g$. Presumably, $\kappa_h \leq{\cal
O}(\kappa_g) $.

Varying the action 
$S=\int d^4 x\sqrt{-g}L$
with respect to
$g_{\mu\nu}$, under fixed $g_h$, we arrive at the unimodular
metagravity equations as follows
\be\label{mgeq}
G_{\mu\nu}\equiv R_{\mu\nu}-\frac{1}{2}Rg_{\mu\nu}=
\frac{1}{\kappa_g^2} T_{\mu\nu}, 
\ \ \  T_{\mu\nu} = T_{ \Lambda\mu\nu}+T_{m \mu\nu}   +T_{h \mu\nu} ,
\ee
with $R_{\mu\nu}$ being the Ricci curvature tensor. In the above,
$T_{\Lambda \mu\nu} =\Lambda g_{\mu\nu}$ is
the vacuum contribution to the total energy-momentum tensor
$T_{\mu\nu}$, with $T_{m \mu\nu}$ being the ordinary  matter
contribution  and  $T_{h \mu\nu}$ the  graviscalar one. 
The latter looks like
\be\label{Th}
T_{h\mu\nu}=\partial_\mu \chi \partial_\nu \chi -\bigg(\frac{1}{2}
\partial\chi \cdot \partial\chi -\check V_h\bigg) g_{\mu\nu},
\ee
where 
\be
\check V_h  =V_h + \kappa_h \Big(\partial V_h/\partial \chi+
\nabla\cdot\nabla \chi\Big),
\ee
with $\nabla_\mu$ standing for a covariant derivative. The contracted
Bianchi identity, $\nabla_\mu G^\mu_\nu=0$,  results in the covariant
conservation of the total energy-momentum, $\nabla_\mu T^\mu_\nu=0$,
instead of $\nabla_\mu  T_m{}^\mu_\nu=0$ for the ordinary matter
alone. (Thereof, the treatment of the graviscalar as DM).
Eq.~(\ref{Th})  resembles that for an ordinary
scalar field in GR except for the metapotential~$\check V_h$
superseding the conventional potential~$V_h$.
When dealing with the metagravity equations, we can
proceed in the canonical coordinates, $g_h=-1$, followed
by a transformation to the  observer's coordinates~$x^\mu$. Instead,
we proceed directly in $x^\mu$, with~$\chi$ taken as an independent
variable, which substitutes a metric element fixed by an additional
coordinate condition. At that,
the unknown $g_h$ does not enter the calculations explicitly. Having
found metric and $\chi$ we can then through Eq.~(\ref{chi})  recover
in the same coordinates the required $g_h$ , solving in a sense an
inverse problem.

In what follows, we restrict ourselves to empty space,
$T_m{}_{\mu\nu}=0$.  In this case,
the contracted Bianchi identity  results in
the graviscalar field equation as follows:$^8$
\be\label{self}
\nabla\cdot\nabla \chi +\partial \check V_h/\partial \chi =0, 
\ee
with $\check V_h$ reduced to  
\be
\check V_h  =V_h - w_h e^{-\chi/\kappa_h}.
\ee
Here,   $w_h $ is an arbitrary integration constant, not a
Lagrangian parameter, which distinguishes the local vacua. 

Particularly, consider the static  spherically symmetric
configuration of metric and the graviscalar field. The line element in
the polar coordinates $(t,r,\theta, \varphi)$ looks generally like
\be\label{polc} 
ds^2= a d  t^2- b d   r^2-  c  r^2 d\Omega, \ \ \   
d\Omega = d \theta^2+\sin^2\theta d\varphi^2,
\ee 
with the three metric potentials $a$, $b$ and $c$ depending on the
radial coordinate $r$ alone. The same is supposed about $\chi$.
Specify $r$ by the coordinate condition $ab=1$ and
choose $\chi$ as the
third independent variable instead of $b$. Neglect by the  potential
$V_h$ and the cosmological constant $\Lambda$. Putting
$X=\chi/\kappa_h$, $A=a=1/b$ and $C= r^2 c$, 
we get  Eq.~(\ref{self}) as
\be\label{X}
(ACX')'=\frac{w_h}{\kappa_h^2} C e^{-X},
\ee
with a prime meaning a derivative with respect to $r$,
and the unimodular metagravity equations in the  vacuum as$^8$
\bea
\label{AC1}
(CA')'&=& \frac{2w_h}{\kappa_g^2}C e^{-X},\\
\label{AC2} 
 (C C')'-\frac{3}{2}C'^2
&=&-\frac{\kappa_h^2}{\kappa_g^2}   (C X')^2,\\
\label{AC3}
(CA')' - (AC')' + 2&=& 0.  
\eea
By construction, Eqs.~(\ref{AC1}) - (\ref{AC3}) are independent,
with  Eq.~(\ref{X}) being an identity. Instead, we choose
Eqs.~(\ref{X}) -- (\ref{AC2}) as the independent ones, with
Eq.~(\ref{AC3}) serving as a constraint.

\paragraph{Anomalous vacuum solution}

Let first $\kappa_h$ be arbitrary, $\kappa_h\leq {\cal O}(\kappa_g)$.
At $w_h=0$, an exact solution to the metagravity equations was given
in Ref.~8. At $r=0$, the solution is singular (reflecting a center
point-like matter). This case  corresponds to
GR in the presence of a scalar field. With $w_h\neq 0$, there
appears a solution regular at $r=0$. Expanding the unknown
functions as the power series in $r $ and equating coefficients at
equal powers on both sides of equations  we have up to terms~$r^6$:
\bea
\label{solX}
X&=&  \tau^2
-\frac{1}{2}\Big(\frac{3}{5}+\varepsilon_h^2\Big)\tau^4
+\bigg(\frac{1}{35}\Big(4+\frac{41}{3}\varepsilon_h^2\Big) +
\frac{1}{3}\varepsilon_h^4\bigg)\tau^6,\\
\label{solA}
a-1 &=&\varepsilon_h^2 \bigg(\tau^2 - \frac{3}{10}\tau^4
+\frac{1}{35}\Big(4+\frac{19}{6} \varepsilon_h^2 \Big)\tau^6\bigg),\\
\label{solC}
c- 1&=& \varepsilon_h^2 \bigg(-\frac{1}{10} \tau^4
+\frac{2}{7}\Big(\frac{1}{5}+\frac{1}{3}
\varepsilon_h^2\Big)\tau^6\bigg),
\eea
where $\varepsilon_h^2 =2\kappa_h^2/\kappa_g^2$ and
$\tau^2=r^2/R_h^2$, with
$R_h^2=6\kappa_h^2/ w_h$ presenting a characteristic length
scale squared. Eq.~(\ref{AC3}) is fulfilled identically up to terms
$\tau^6$.   

Continuing the procedure above we can  find the solution
with any desired accuracy. Namely, knowing the solution 
in an order $\tau^{2n}$, $n=0,1,2,\dots$ we can first
determine $X$ from the l.h.s.\ of Eq.~(\ref{X}) in the next order
$\tau^{2(n+1)}$. Then we can find  $a$ and $c$ in the same order from
Eqs.~(\ref{AC1}) and (\ref{AC2}), respectively, {\it etc}. At that,
$X(0)$, $a(0)$ and $c(0)$ are fixed by the boundary conditions at
$\tau=0$. Both  $w_h>0$  and $w_h<0$ 
are {\it a priori} envisaged. The respective solutions are formally
related by substitution $\tau^2\to -\tau^2$. 
For physical reasons, $w_h\geq 0$ (see later).  

Of special interest is the case $\varepsilon_h\ll 1$. Decomposing
an exact solution, regular at the center, as the power series in
$\varepsilon_h$ (only even powers enter) as $X=\sum
\varepsilon_h^{2n}X_n$, $a= \sum \varepsilon_h^{2n}a_n$ and $c=\sum
\varepsilon_h^{2n}c_n $,
$n=0,1,\dots$, with $a_0=c_0=1$, we simplify the
metagravity equations in the appropriate leading orders as follows: 
\bea 
\label{leadX}
\frac{ d}{d \tau} \bigg(\tau^2 \frac{d X_0}{d \tau}\bigg)&=& 6 \tau^2
e^{-X_0},\\
\label{leada}
\frac{ d}{d \tau} \bigg(\tau^2 \frac{d a_1}{d \tau}\bigg)&=& 6
\tau^2 e^{-X_0} ,\\
\label{leadc}
\frac{ d}{d \tau} \bigg(\tau^2 \frac{d c_1}{d \tau}\bigg)
&=&-\frac{1}{2}\bigg(\tau \frac{ d X_0}{d\tau}\bigg)^2,
\eea
with the restriction
\be\label{leada-c}
\frac{d}{d \tau}\bigg(\tau^2  \frac{d} {d \tau} (a_1-c_1)-
2 \tau(a_1+c_1)\bigg)= 0.
\ee\label{leadac}
Clearly, it is possible to add to the solutions for  $a_1$ and
$c_1 $ the arbitrary reciprocal
terms $\sim1/\tau$. Assuming no singularity in the center, we 
omit such contributions. It follows from the equations above that the
driving term in the system is~$X_0$.
Having found the latter in a self-consistent manner from
Eq.~(\ref{leadX}) we 
can then find $a_1$ and $c_1$ from the two other equations with an
external source determined by $X_0$. In particular, it follows that
$a_1 =X_0$ modulo a constant which may be put to zero. The
leading in $\varepsilon_h$ part of the regular solution given by 
Eqs.~(\ref{solX}) -- (\ref{solC}) explicitly
satisfies all these equations up to accuracy~$\tau^6$. 

To  study $X_0$ at $\tau^2\ge 0$ {\it in toto}
note first of all that there exists an exact exceptional solution of
Eqs.~(\ref{leadX}) -- (\ref{leadc}) as follows:
\bea
\label{partX}
\bar X_0 &=&\ln 3\tau^2,\\
\bar a_1 &=&\ln 3\tau^2 ,\\
\bar c_1 &=&-\ln 3\tau^2  + 2,
\eea
with  the additive constants restricted by the relation $\bar a_1
=\bar X_0$ and Eq.~(\ref{leada-c}). 
Present further Eq.~(\ref{leadX}) as follows:
\be\label{ddY}
\frac{d^2 Z}{d \si^2}+\frac{1}{2} \frac{d Z}{d \si}=\frac{1}{2}
(e^{-Z}-1)
\ee
where $Z= X_0-\si$, with  $\si=\ln 3\tau^2$ any real,
$-\infty<\si<+\infty$. 
Introducing  $\dot Z \equiv d Z/d \si$ as an independent variable
supplementing $Z$, reduce the second-order Eq.~(\ref{ddY}) to the
equivalent  autonomous first-order system:
\bea
\frac{d Z}{d \si}&=& \dot Z,\\
\frac{d \dot Z}{d \si} &=&-\frac{1}{2} \dot
Z+\frac{1}{2}(e^{-Z}-1).
\eea
In the
phase plane $(Z,\dot Z)$, there is  a single exceptional point $\bar
Z=\bar {\dot Z}=0$, defined by the requirement $d Z/d \si=d \dot Z/d
\si=0$,
other points being normal. Through each normal point
there should come precisely one phase trajectory $(Z(\si),\dot
Z(\si))$.
The latter ones satisfy the equation
\be\label{iso}
\frac{d \dot Z}{d Z}= \frac{1}{2 \dot Z}(e^{-Z}-1)-\frac{1}{2},
\ee
with the isoclines $d \dot Z/d Z=m$ being
\be
\dot Z=\frac{1}{2m +1}(e^{-Z}-1),
\ee
where $m$ is an arbitrary constant. At that, the axes $\dot Z=0$ and
$Z= 0$  correspond to  $m\to\pm \infty$  and $m=-1/2$, respectively. 

Inspection of the phase plane shows that the exceptional  point in the
center belongs to the stable focus type, with all
the trajectories winding round the center and approaching
the latter with $\si\to+\infty$. At that, the exceptional point
presents the exceptional solution Eq.~(\ref{partX}). 
There is a unique trajectory
with $\dot Z$ remaining finite at  $\si\to -\infty$, namely, 
$\dot Z\to -1$, and behaving thus
like $Z\simeq -\si$ asymptotically. 
Such a trajectory 
corresponds to the regular at $\tau=0$ solution $X_0$ given by
Eq.~(\ref{solX}) with $\varepsilon_h=0$. The rest of
trajectories satisfy $\dot Z\to - \infty$ at $\si\to -\infty$,
with the respective $X_0$ being thus irregular at $\tau=0$.
The regular anomalous solution  is  stable against small perturbations
of the initial data taken on the axis $\dot Z=0$, $ Z<0$, but for very
small $\tau^2>0$.  In the latter region,  the regular anomalous
solution is
to be superseded by  a  singular normal solution with $r_h\ll R_h$,
where $r_h$ is the graviscalar radius of a center singularity.$^8$ The
account for the latter does not significantly affect the halo 
at $\tau\gg r_h/R_h$.  

Altogether, the regular anomalous vacuum solution  for the graviscalar
field $X$ at $\varepsilon_h\ll 1$ looks like:
\be\label{X0}
X_0= \cases{\tau^2 -\frac{3}{10}\tau^4
+\frac{4}{35}\tau^6 
+{\cal O}(\tau^8),\ \  \mbox{\rm at}\ \ 0\leq\tau < 1,
\cr
\ln 3 \tau^2, \hspace{15ex} \ \ \ \ \ \ \ \ 
\mbox{\rm at}\ \  \tau\gg 1.  }
\ee
It oscillates around the exceptional solution $\bar X_0$
approaching the latter at $\tau\gg 1$.

\section*{3. Anomalous Rotation Curves}

\paragraph{Graviscalar DM}

The velocity of circular rotation of a test particle in the static
spherically symmetric metric Eq.~(\ref{polc}) is given~by
\be\label{v2}
v^2=\frac{a'}{(\ln r^2 c )'}.
\ee
So defined velocity transforms as a scalar under the local radial
transformations. To get~$v^2$ in the leading  $\varepsilon_h$-order
we can put $c=c_0=1$.  At $w_h> 0$, the regular anomalous solution
results in  the scaled RC profile as follows:
\be\label{vh2'}
v_{h}^2(\tau)= \frac{\varepsilon_h^2}{2} \frac{\tau d a_1}{d \tau}=
\varepsilon_h^2\cases{\tau^2 -\frac{3}{5}\tau^4 
+\frac{12}{35} \tau^6 
+ {\cal O}(\tau^8),
\ \  \mbox{\rm at}\ \  0\leq\tau < 1,\cr
1, \hspace{20ex} \ \ \ \ \ \    \mbox{\rm at}\ \ \tau \gg 1,}
\ee
where  use is made of $a_1=X_0$.
At that, the exceptional solution results in the flat~RC
\be
\bar v_{h}^2(r)=\varepsilon_h^2,  
\ee
around which  all the  RC's $v_{h}^2(r)$, with
different $R_h$, oscillate approaching $\bar v_{h}^2$ at $r\gg R_h$.

Let us now interpret RC's  in terms of DM. 
The Newton dynamics in flat space ($a=c=1$) with a DM would result in 
\be\label{N}
\frac{v_h^2}{r}=\frac{ G  M_h(r)}{r^2},
\ee
where $ M_h(r) =4\pi\int_0^r  \rho_h(r) r^2d r$ is
the DM energy interior to $r$, with $ \rho_h$ being the DM 
energy density. This implies
\be
\rho_h=\frac{1}{4\pi G} \frac{(r
v_h^2)'}{r^2}.
\ee
To reproduce the first part of Eq.~(\ref{vh2'}) we should have 
\be
\rho_{h}=
\frac{\varepsilon_h^2\kappa_g^2}{R_h^2}\frac{1}{\tau^2}\frac{d}{d
\tau}
\bigg(\tau^2\frac{d a_1}{d \tau}\bigg).
\ee
(For a center  point-like matter with $a-1\sim -1/r$ and 
$v^2\sim 1/r$, this would  give $\rho_h =0$.) 
Accounting for Eq.~(\ref{leada}),   
we get finally the looked-for  DM profile  as follows ($\tau^2\ge 0$): 
\be\label{DM}
\rho_{h}(\tau) = 2w_h e^{-X_0}=\rho_h(0)   \cases{1-\tau^2
+\frac{4}{5}\tau^4  
+{\cal O}(\tau^6), \ \ \mbox{\rm at}\ \  0\leq\tau< 1, \cr
1/(3\tau^2),  \hspace{15ex}\,  \mbox{\rm at}\ \  \tau\gg 1, }
\ee
with the central density
\be\label{DM1}
\rho_h(0)=\frac{6\varepsilon_h^2 \kappa_g^2}{R_h^2}.
\ee
Asymptotically,  $ M_{h}(r) \simeq
\varepsilon_h^2 r/G$. Ultimately, such a linear growth should be
terminated by the potential $V_h$, which would become significant at
the periphery, where $X_0$ gets strong. 
Thus the regular anomalous solution corresponds to a smooth DM halo
with the finite central density.
At that, the exceptional solution results in the cuspy profile
\be
\bar\rho_{h}(r)
=\frac{2\varepsilon_h^2\kappa_g^2}{r^2},
\ee
with the  exact $\bar M_{h}(r) =\varepsilon_h^2 r/G$.  The family of
the smooth profiles
$\rho_{h}(r)$, with various $R_h$, oscillates  around $\bar\rho_{h}$ 
approaching the latter at $r\gg R_h$. 

According to Ref.~8, $\rho_h=-2\check V_h$ may be treated as  the
energy density  of a static graviscalar field, incorporating its
gravitational energy. This insures a dual field-matter interpretation
of the graviscalar halo. In terms of field, the case $w_h>0$  presents
a local vacuum well, with the metapotential $\check V_h= -w_h
e^{-X_0}$ due to a
coherent state  $X_0$ of the graviscalar  field. In terms of matter,
the same case corresponds to the DM distribution 
with   $\rho_h>0$ which produces in  flat space precisely the same
attraction.  The case $w_h<0$ 
presents a local vacuum bump with the repulsive $\rho_h<0$ implying 
an unstable configuration ($v_h^2<0$). In the
GR limit, $w_h=0$, the halo clearly disappears.

\paragraph{Galaxy halos}

There are numerous studies in astrophysical literature concerning
the galaxy DM halos. At that, the empirical  halo density profiles
rely mostly on
the two-component fits to the galaxy RC's, with the matter and halo
contributions  added in quadrature, $v^2=v_m^2+v_h^2$, where  $v_m^2$
is a total contribution  of the different types of luminous matter
(disk, gas,  bulge) and $v_h^2$ is a halo contribution. Thereof,
there emerge ever growing evidences, based on a vast sample of
galaxies of different types, in favour of the DM halos 
with the finite central density profiles
 (see, {\it e.g.}, Refs.~9 -- 11, with
an extensive list of  references therein).  In particular,
in Ref.~10 it is found a universal DM density profile,
extracted from a sample of 36 nearby spiral galaxies, as follows:
\be\label{halo}
\rho_h=\frac{\rho_0}{1+(r/R_0)^2}
\ee
with $\rho_0$ being a central density and $R_0$ a core radius.
The empirical smoothness displayed by Eq.~(\ref{halo}) is at
sharp variance with the cuspy form of the cold DM halos. In contrast,
the vacuum
graviscalar halo obtained in the present paper  naturally complies
with smoothness.  Moreover, Eq.~(\ref{DM}) closely reproduces 
the first three terms of the decomposition of Eq.~(\ref{halo}).
Nevertheless, there are two differences. First,  Eq.~(\ref{DM1})
implies $\rho_{h}(0)\sim R_h^{-2} $, whereas empirically
there emerges a constant central surface density of the galaxy halos,
{\it i.e.}, $\rho_0 R_0\sim {\rm const} $. Second, Eq.~(\ref{DM}) is
three times lower asymptotically compared to  Eq.~(\ref{halo}).

The reason of the discrepancy may be as follows.  
We have restricted ourselves by the simplest model with the spherical
graviscalar halo in empty space. Such a halo may serve just as a
prototype for the real galaxies. To confront the theory with the data
the luminous matter should also be accounted for. This would give 
$v^2=v_m^2+  v_{h m}^2 $, with $ v_{h m}^2 $ being
the effective graviscalar halo contribution in the presence of matter.
Because of a coherent nature of  halo the deformation of the latter,
both in magnitude and sphericity, may be significant within the region
of intersection of matter and halo. It is rather $v_{h m}^2$, to
which the two-component fit Eq.~(\ref{halo}) is to be applied,
than~$v_{h }^2$ due the vacuum graviscalar halo. To
distil the galaxy RC sample from the matter contribution as far as
possible, live aside
the points from Ref.~10 which correspond explicitly to the luminous
matter dominance at the distances at hand. Inspection shows that
there are at least four, out of 36, points on the $\log \rho_0$ --
$\log R_0$ plot, with the extremely large $R_0$, to be dropped off. 
(Incidentally, such peculiar $R_0$ are strongly model dependent.) The
rest of points lies much more compactly,  the subsequent results
being less sensitive to further reducing the galaxy sample. On the
reduced sample, the dependence $\log \rho_0\sim -\log R_0$ looks less
prominent. On the other hand, a much wider sample of galaxies of
different types still supports the constant central surface density
law.$^{11,}$\footnote[2]{The latter data may also indicate  some
deviations from Eq.~(\ref{halo}) both at very small and very large
galactocentric distances.}  To settle the question  in the metagravity
framework the account for the  luminous matter, 
with an axisymmetric
distribution resulting in  the graviscalar halo asphericity, 
is required.

Finally note that it is the Lagrangian parameter
$\varepsilon_h=\sqrt{2}\kappa_h/\kappa_g$, which sets the scale of the
asymptotically flat RC's due to the graviscalar halo in the vacuum.
So, taking for galaxies asymptotically 
$v_h(\infty)\sim 100$~km/s  we would expect that 
$\varepsilon_h= v_h(\infty)/c \sim 10^{-3} $. With $\kappa_g=2.4
\times 10^{18}$~GeV,  the unimodular metagravity mass scale, 
$\kappa_h \sim 10^{15} $~GeV, would
approach the GUT mass scale.

\section*{4. Conclusion}

The regular anomalous solution to the static spherically symmetric 
metagravity equations in empty space presents a viable prototype model
for the smooth galaxy halos characterized by the finite central
density profiles. It goes without saying that once the solution models
the descent density profiles of the DM halos, it provides to the same
extent all the other
gravitational effects of such halos.  The hypothesis about the
graviscalar origin of DM in the framework of the unimodular
metagravity  finds thus its preliminary confirmation. To further
verify the theory studying the graviscalar halos in
the presence of the axisymmetric matter distribution is in order.

\section*{References}

1. J.J.~van der Bij, H.~van Dam,  and Y.J.~Ng, {\it Physica} {\bf
116A}, 307 (1982).\\
2. J.L.\ Anderson and D.R.~Finkelstein, {\it Am.\ J.\ Phys.} 
{\bf 39}, 901 (1971). \\
3. D.R.~Finkelstein, A.A.~Galiautdinov, and J.E.~Baugh, {\it J.\
Math.\ Phys.} {\bf 42}, 340 (2001), gr-qc/0009099.\\ 
4. L.~Smolin, arXiv:0904.4841 [gr-qc].\\
5. M.~Shaposhnikov and D.~Zenhausern, {\it Phys.\ Lett.\  B} 
   {\bf 671}, 187 (2009), arXiv:0809.3395 [hep-th].\\
6. W.~Buchm\"uller and N.~Dragon, {\it  Phys.\ Lett.\  B} {\bf 207},
292 (1988). \\
7. Yu.F.~Pirogov, {\it Phys.\ At.\ Nucl.}  {\bf 69}, 1338 (2006),
   gr-qc/0505031.\\
8. Yu.F.~Pirogov, arXiv:0903.2018 [gr-qc].\\
9. M.~Persic, P.~Salucci and F.~Stel,  {\it MNRAS} {\bf281}, 27
(1996), astro-ph/9506004.\\ 
10. M.~Spano {\it et al},  {\it MNRAS} {\bf 383}, 297 (2008),
arXiv:0710.1345[astro-ph].\\
11. F.~Donato {\it et al}, arXiv:0904.4054[astro-ph.CO].\\

\end{document}